\documentclass[aps,twocolumn,showpacs]{revtex4}
\usepackage{amsfonts,amsmath,amssymb,epsfig}

\newcommand{\crp}{{\mathcal{N}}}
\newcommand{\bldn}{{\mathrm{\bf n}}}
\newcommand{\heff}{\hbar_{\mathrm{eff}}}
\newcommand{\vecq}{{\bf q}}
\newcommand{\vecp}{{\bf p}}
\newcommand{\vecz}{{\bf z}}
\newcommand{\vecx}{{\bf x}}

\newcommand{\rem}[1]{}

\newcommand{\ui}{\mathrm{i}}
\newcommand{\ue}{\mathrm{e}}

\begin{document}

\title{The Quantum Normal Form Approach to Reactive Scattering: The Cumulative Reaction Probability for Collinear Exchange Reactions}

\author{Arseni Goussev$^1$, Roman Schubert$^1$,  Holger Waalkens$^{1,2}$, and Stephen Wiggins$^1$}

\affiliation{$^1$School of Mathematics, University of Bristol, University Walk, Bristol BS8 1TW, UK\\
$^2$Department of Mathematics and Computing Science, University of Groningen, PO Box 407,  9700 AK Groningen, The Netherlands}

\date{\today}

\begin{abstract}
  The quantum normal form approach to quantum transition
  state theory is used to compute the cumulative reaction
  probability for collinear exchange reactions.
  It is shown that for heavy atom systems like the nitrogen exchange reaction 
  the quantum normal form approach gives excellent results and has major 
  computational benefits over full reactive scattering approaches. For light atom systems 
  like the hydrogen exchange reaction however the quantum normal approach is shown to give only poor results.  
  This failure  is attributed  to the importance of tunnelling trajectories in light atom reactions
  that are not captured by the quantum normal form as indicated by the only very slow convergence 
  of the quantum normal form for such systems. 
\end{abstract}

\pacs{34.10.+x, 34.50.Lf, 05., 02.70.-c, 02.,03.65.Xp,82.20.-w,82.20.Db,82.20.Ej,82.30.Hk}

\maketitle

\section{Introduction}

The classical mechanical picture of a chemical reaction as a
scattering problem across a saddle point of the Born-Oppenheimer
potential energy surface in configuration space has proven to be a
fruitful way of visualizing and thinking about chemical reactions
since the 1930's, when Eyring, Polanyi, and Wigner developed {\em
  transition state theory} (TST). TST provides the framework for
computing, using classical mechanics, many of the physically important
quantities for describing such chemical reactions. The fundamental
geometrical object in TST is a dividing surface that divides the
energy surface into a reactant and a product component. With such a
dividing surface in hand, one can then compute the reaction rate from
the directional phase space flux through this surface.  In order not
to overestimate the rate the dividing surface must not be recrossed by
reactive trajectories, i.e. the dividing surface should have the ``no
re-crossing'' property.  In the 70's Pechukas, Pollak and others
\cite{PechukasMcLafferty73,PechukasPollak78} showed that for two
degrees of freedom such a dividing surface can be constructed from a
periodic orbit (the so called \emph{periodic orbit dividing
  surface}). Recently it has been shown that for more than two
degrees-of-freedom a dividing surface that is free of recrossings can
be built from a {\em normally hyperbolic invariant manifold} (NHIM)
\cite{Wiggins94}. The dividing surface and the NHIM can be directly
constructed from an algorithm based on a Poincar{\'e}-Birkhoff normal
form procedure \cite{UJPYW01} which also gives an expression for the
flux \cite{WaalkensWiggins04}. The classical phase space transition
state theory, based on Poincar\'e-Birkhoff normal form theory,
naturally leads to a quantum version of transition state theory, based
on a quantum normal form. Since the normal form is valid in a
neighborhood in energy both above and {\em below} the saddle point, it
includes the quantum effect of tunneling in the region near the saddle.  
Moreover, it does not
require a full quantum simulation in a neighborhood of the TST
dividing surface (\cite{schubert_prl,wsw08}) in order to compute
important quantities associated with the reaction. This is significant
since much effort has been devoted to developing a quantum version of
transition state theory whose implementation remains feasible for
multi-dimensional systems (see the flux-flux autocorrelation function
formalism by Miller and coworkers \cite{Miller98}). However, in
\cite{Miller1} Miller stated that ``--the conclusion of it all is that
there is no uniquely well defined quantum version of TST in the sense
that there is in classical mechanics. This is because tunneling along
the reaction coordinate necessarily requires one to solve the
(quantum) dynamics for some finite region about the TS dividing
surface, and if one does this quantum mechanically there is no
`theory' left, {\em i.e.}, one has a full dimensional quantum dynamics
treatment that is {\em ipso facto} exact, a quantum simulation.''
Nevertheless, our approach based on the quantum normal form leads to a
quantum version of transition state theory that includes tunneling near the saddle and
does not require a full quantum simulation in a neighborhood of the
TST dividing surface.  Moreover, our computation of the cumulative
reaction probability can be viewed as the quantum mechanical flux
through a (classically recrossing free) dividing surface, which
includes tunneling. The quantum normal form gives a local decoupling
of the quantum dynamics to any desired order in $\hbar$ which is the
key issue here, i.e. locally, we have a decoupling of the scattering
states into forward/backward reactive and non-reactive, and for these states we know the
transmission probabilities analytically. Therefore we do not have to
`simulate' the quantum dynamics. Hence, it 
sidesteps the
issues and concerns expressed by Miller.

In this paper we illustrate the utility of the quantum normal form
approach to quantum transition state theory by considering the
computation and behavior of the bimolecular {\it cumulative reaction
  probability} (CRP) $\crp(E)$, defined as \cite{seideman,miller98}

\begin{equation}
  \crp(E) = \sum_{\bldn_\mathrm{r},\bldn_\mathrm{p}} |S_{\bldn_\mathrm{r},\bldn_\mathrm{p}}(E)|^2,
\label{1-01}
\end{equation}

\noindent
where $S(E)$ is the reactive scattering matrix evaluated at energy
$E$, and $\bldn_\mathrm{r}$ ($\bldn_\mathrm{p}$) are the quantum
numbers describing the asymptotic channel of incoming reactants
(outgoing products). The CRP is a fundamental quantity that
characterizes the reaction rate: the microcanonical and canonical rate
constants can be determined from $\crp(E)$ by means of simple
relations \cite{seideman}.

This paper is outlined as follows. In Section \ref{sec_theory} we
outline the theoretical and computational aspects of the quantum
normal form theory. We emphasize the structural features that allow
for the treatment of high dimensional quantum problems and show how
the quantum normal form leads to a ``simple'' expression for the
CRP. In Section \ref{sec_reactions} we apply the quantum normal form
approach to the computation of the CRP for the collinear hydrogen and
nitrogen exchange reactions. These quantities are compared to the
``exact'' answer obtained from a reactive quantum scattering
calculation. In Section \ref{sec_convergence} we discuss some aspects
related to the convergence properties of the quantum normal form, and
in Section \ref{sec_conclusions} we summarize our results and offer
some directions for further investigations.


\section{Quantum normal form theory}
\label{sec_theory}

In this section we present central aspects of the  quantum normal form (QNF) theory; for rigorous
mathematical statements, proofs and further details we refer to
Ref.~\cite{schubert_prl,wsw08}.

We begin by considering a quantum Hamilton operator $\hat{H}$ which we
assume to be obtained from the Weyl quantization of a classical
Hamilton function $H(\vecp,\vecq)$. Here $\vecq = (q_1, q_2, \ldots,
q_d)$ and $\vecp = (p_1, p_2, \ldots, p_d)$ denote the canonical
coordinates and momenta, respectively, of a Hamiltonian system with
$d$ degrees of freedom.  Throughout this paper we will use atomic
units, so that $\vecq$ and $\vecp$ are dimensionless.  We will denote
the corresponding operators by $\hat{\vecq} = (\hat{q}_1, \hat{q}_2,
\ldots, \hat{q}_d)$ and $\hat{\vecp} = (\hat{p}_1, \hat{p}_2, \ldots,
\hat{p}_d)$. In the coordinate representation their components
correspond to multiplication by $q_j$ and the differential operators
$\hat{p}_j = -\ui \heff \, \partial / \partial q_j$. Here $\heff$ is a
dimensionless parameter which corresponds to a scaled, effective
Planck's constant. For molecular reactions described in the
Born-Oppenheimer approximation, $\heff^2$ occurs naturally as the
ratio of the electronic mass and the reduced mass of the nuclei
participating in the reaction as we will see below in more detail.

The main idea of the QNF procedure is to approximate the Hamilton
operator $\hat{H}$ by a simpler Hamilton operator obtained from a
power series expansion of $\hat{H}$ which is simplified order by order
using unitary transformations.  As we will describe in more detail in
Sec.~\ref{sec_convergence} the scaled Planck's constant, $\heff$, will
play the role of a `small parameter' which controls the quality of the
QNF approximation.  For our application of bimolecular reactions, the
resulting transformed Hamilton operator truncated at a suitable order
will be simpler in the sense that it will provide an easy, explicit
way to compute the cumulative reaction probability.

To define and implement the unitary transformations it is extremely
beneficial not to work with operators but with their Weyl
\emph{symbols} instead.  The Weyl symbol of an operator $\hat{H}$ is
defined as
\begin{equation}
  H^{(0)}(\vecq,\vecp;\heff) = \int \! d\vecx \, \langle \vecq-\vecx/2 |
  \hat{H} | \vecq+\vecx/2 \rangle \, e^{i \vecp \vecx / \heff} .
\label{2-01}
\end{equation}
The superscript $(0)$ is introduced for reasons that will become clear
in a moment.  The map $\hat{H} \mapsto H^{(0)}(\vecq,\vecp;\heff)$
leading to \eqref{2-01} is also called the Wigner map. It is the
inverse of the transformation which yields a Hamilton operator
$\hat{H}$ from the Weyl quantization, $\mathrm{Op}[H] $, of a phase
space function $H$ (the Weyl map) which, using Dirac notation, is
given by
\begin{equation}
\begin{split}
\hat{H}=  \mathrm{Op}[H] =& \iint \frac{d\vecq d\vecp}{\left( 2\pi\heff \right)^d}
  H(\vecq,\vecp)  \\ & \times \displaystyle \int
  d\vecx | \vecq-\vecx/2 \rangle e^{-i\vecp\vecx / \heff}
  \langle \vecq+\vecx/2 | \,.  
\end{split} \label{eq:def_Weyl_map}
\end{equation}
Accordingly, $H^{(0)}(\vecq,\vecp;\heff)$ in \eqref{2-01} agrees with
the classical Hamilton function $H(\vecq,\vecp)$ in our case. The
argument $\heff$ is introduced for convenience since the Weyl symbol
of the unitarily transformed Hamilton operator will in general
explicitly depend on $\heff$.

We will now assume that $H^{(0)}(\vecq,\vecp;\heff)$ (or equivalently
$H(\vecq,\vecp)$) has a (single) equilibrium point, $\vecz_0 \equiv
(\vecq_0, \vecp_0)$, of saddle-center-$\ldots$-center stability
type. By this we mean that the matrix associated with the
  linearization of Hamilton's equations about this equilibrium point
  has two real eigenvalues, $\pm \lambda$, of equal magnitude and opposite sign,  
  and $d-1$ purely imaginary complex conjugate pairs of
  eigenvalues $\pm \ui \omega_k$, $k=2,\ldots,d$. 
  If the classical Hamiltonian
  is of the form kinetic energy plus potential energy then these type
  of equilibrium points of Hamilton's equations correspond to index
  one saddle points of the potential energy. 
%
%
Using the symbol calculus the QNF theory provides a systematic
procedure to obtain a local approximation, $\hat{H}_{\mathrm{QNF}}$,
of the Hamiltonian $\hat{H}$ in a phase-space neighborhood of the
equilibrium point $\vecz_0$ in order to facilitate further computation
of various quantities, such as the CRP, of the reaction system under
consideration. In the following we summarize the essential steps of
the QNF procedure.

The QNF procedure consists of a sequence of, in general $\heff$ dependent,
generalized phase-space coordinate transformations changing the
symbol as
\begin{equation}
H^{(0)} \rightarrow H^{(1)} \rightarrow H^{(2)} \rightarrow H^{(3)}
\rightarrow \ldots \rightarrow H^{(N)}. 
\label{eq:seq_trafos}
\end{equation}
The first of the transformations \eqref{eq:seq_trafos}
shifts the equilibrium point
$\vecz_0$ to the origin according to 
\begin{equation}
  H^{(1)}(\vecz;\heff) = H^{(0)}(\vecz+\vecz_0;\heff) \, ,
\label{2-02}
\end{equation}
where $\vecz \equiv (\vecq, \vecp)$. Once the equilibrium point is
shifted to the origin, the QNF procedure deals with the Taylor expansion of
the symbols in $\vecz$ and $\heff$:
\begin{equation}
  H^{(n)}(\vecz;\heff) = E_0 + \sum_{s=2}^\infty H_s^{(n)}(\vecz;\heff) \, ,
\label{2-03}
\end{equation}
with
\begin{equation}
\begin{split}
  H_s^{(n)}(\vecz;\heff)  = & \displaystyle \sum_{|\alpha|+|\beta|+2j=s}
    \frac{H_{\alpha_1,\ldots,\alpha_d,\beta_1,\ldots,\beta_d,j}^{(n)}}{\alpha_{1}!
      \ldots \alpha_{d}! \beta_{1}! \ldots \beta_{d}! j!}  \\ 
    &\times q_1^{\alpha_1} \ldots q_d^{\alpha_d} p_1^{\beta_1} \ldots
    p_d^{\beta_d} \, \heff^{j} \, ,
\end{split} \label{2-04}
\end{equation}
where $\alpha_k,\beta_k,j \in \mathbb{N}_0$, $|\alpha| = \sum_k
\alpha_k$, $|\beta| = \sum_k \beta_k$, and
\begin{eqnarray}
  \lefteqn{ H_{\alpha_1,\ldots,\alpha_d,\beta_1,\ldots,\beta_d,j}^{(n)} }
  \nonumber\\ &= \displaystyle \left.
  \prod_{k,l=1}^d \frac{\partial^{\alpha_k}}{\partial
    q_k^{\alpha_k}} \frac{\partial^{\beta_l}}{\partial
    p_l^{\beta_l}} \frac{\partial^{j}}{\partial \varepsilon^{j}}
  H^{(n)}(\vecz; \varepsilon) \right|_{({\bf 0};0)} \!\!\! .
  \label{2-04.1}
\end{eqnarray}

At the next step of the transformation sequence one finds a symplectic $2d
\times 2d$ matrix $M$ such that the second order term of the symbol
\begin{equation}
  H^{(2)}(\vecz; \heff) = H^{(1)}(M^{-1} \vecz; \heff)
\label{2-04.2}
\end{equation}
takes the particularly simple form:
\begin{equation}
  H_2^{(2)}(\vecz; \heff) = \lambda q_1 p_1 + \sum_{k=2}^d \frac{\omega_k}{2}
  (q_k^2 + p_k^2) \,.
\label{2-05}
\end{equation}
Section 2.3 of Ref.~\cite{wsw08} provides an explicit procedure for
constructing the transformation matrix $M$.

In order to proceed with the higher order transformations of the symbol of the
Hamiltonian it is essential to introduce the notion of the {\it Moyal
  bracket}. Given two symbols $A(\vecz;\heff)$ and $B(\vecz;\heff)$,
corresponding to operators $\hat{A}$ and $\hat{B}$ respectively, the Moyal
bracket
\begin{equation}
\begin{split}
 \{ A, B \}_{\mathrm{M}}  =  \frac{2}{\heff} A
\sin\left[ \frac{\heff}{2} \sum_{j=1}^d \left( \frac{\overleftarrow{\partial}}{\partial q_j}
    \frac{\overrightarrow{\partial}}{\partial p_j} - \frac{\overleftarrow{\partial}}{\partial p_j}
    \frac{\overrightarrow{\partial}}{\partial q_j} \right) \right] B
\end{split} \label{2-06}
\end{equation}
gives the Weyl symbol of the operator $i[\hat{A},\hat{B}]/\heff$, where
$[\cdot,\cdot]$ denotes the commutator. The arrows in \eqref{2-06}
indicate whether the
partial differentiation acts to the left (on $A$) or to the right (on $B$).
Equation~(\ref{2-06}) implies that, in general, for $\heff \rightarrow 0$,
\begin{equation}
  \{ A, B \}_{\mathrm{M}} = \{ A, B \} + \mathcal{O}(\heff^2) \, ,
\label{2-07}
\end{equation}
where $\{\cdot,\cdot\}$ denotes the Poisson bracket. Moreover, if at least
one of the functions $A$, $B$ is a second order polynomial in the variables
$\vecq$, $\vecp$ then $\{ A, B \}_{\mathrm{M}} = \{ A, B \}$. Finally, to
simplify further notations we define the Moyal-adjoint operator as
\begin{equation}
  \mathrm{Mad}_A : B \mapsto \mathrm{Mad}_A B \equiv \{ A, B \}_{\mathrm{M}}
  \, .
\label{2-08}
\end{equation}

Continuing with the sequence of transformations of the symbol in \eqref{eq:seq_trafos}
we define the spaces
\begin{eqnarray}
  \lefteqn{ \mathcal{W}^n = \mathrm{span} \left\{ q_1^{\alpha_1} \ldots q_d^{\alpha_d}
    p_1^{\beta_1} \ldots p_d^{\beta_d} \heff^{j} : \right. } \nonumber\\ &
    \left. \phantom{p_d^{\beta_d}+++++++} |\alpha|+|\beta|+2j = n \right\} \, .
\label{2-09}
\end{eqnarray}
Then, the symbol $H^{(n)}$ with $n \ge 3$ is obtained from $H^{(n-1)}$ by
means of the transformation generated by a function $W_n(\vecz;\heff) \in
\mathcal{W}^n$,
\begin{equation}
  H^{(n)} = \sum_{k=0}^\infty \frac{1}{k!} \left[ \mathrm{Mad}_{W_n} \right]^k
  H^{(n-1)} \, .
\label{2-10}
\end{equation}
The structure of the transformation defined by Eq.~(\ref{2-10}) implies
\cite{wsw08} that the operators $\hat{H}^{(n)}$ and $\hat{H}^{(n-1)}$
corresponding respectively (through the Weyl quantization) to the symbols
$H^{(n)}$ and $H^{(n-1)}$ are related to one another by means of the unitary
transformation $\hat{H}^{(n)} = e^{\ui \hat{W}_n / \heff} \hat{H}^{(n-1)} e^{-\ui
  \hat{W}_n / \heff}$, where $\hat{W}_n$ is the operator corresponding to the
symbol $W_n$. In terms of the Taylor expansion defined in
Eqs.~(\ref{2-03}-\ref{2-04.1}) the transformation introduced by
Eq.~(\ref{2-10}) reads
\begin{equation}
  H^{(n)}_s = \sum_{k=0}^{\left\lfloor \frac{s}{n-2} \right\rfloor}
  \frac{1}{k!} \left[ \mathrm{Mad}_{W_n} \right]^k H^{(n-1)}_{s-k(n-2)} \, ,
\label{2-11}
\end{equation}
where $\lfloor \cdot \rfloor$ gives the integer part of a number, i.e., the
`floor'-function. Using Eq.~(\ref{2-11}) one can show that the transformation
defined by Eq.~(\ref{2-10}) satisfies the following properties for $n \ge 3$:
\begin{equation}
  H_s^{(n)} = H_s^{(n-1)} \, , \;\;\; \mathrm{for} \;\;\; s < n \, ,
\label{2-12}
\end{equation}
so that, in particular, $H_2^{(n)} = H_2^{(2)}$, and
\begin{equation}
  H_n^{(n)} = H_n^{(n-1)} - \mathcal{D} W_n \, ,
\label{2-13}
\end{equation}
where
\begin{equation}
  \mathcal{D} \equiv \mathrm{Mad}_{H_2^{(2)}} = \{ H_2^{(2)} , \cdot \} \: .
\label{2-14}
\end{equation}
Equation~(\ref{2-13}) is referred as to the {\it quantum homological
  equation}.

We now specify the generating function $W_n$ by requiring $\mathcal{D}
H_n^{(n)} = 0$, or equivalently $H_n^{(n)}$ to be in the kernel of the restriction of
$\mathcal{D}$ to $\mathcal{W}^n$; in view of Eq.~(\ref{2-13}) this condition
yields
\begin{equation}
  H_n^{(n-1)} - \mathcal{D} W_n \in \mathrm{Ker} \,\mathcal{D} |_{\mathcal{W}^n} \, .
\label{2-15}
\end{equation}
Section 3.4.1 of Ref.~\cite{wsw08} provides the explicit
procedure of finding the solution of Eq.~(\ref{2-15}).  Provided the
linear frequencies $\omega_2,\ldots,\omega_d$ in \eqref{2-05} are
rationally independent, i.e. $m_2\omega_2+\ldots+m_d \omega_d=0$
implies $ m_2=\ldots=m_d=0$ for all integers $m_2,\ldots,m_d$, it
follows that for odd $n$, $H_n^{(n)} =0$, and for even $n$,
\begin{equation}
  H_n^{(n)} \in \mathrm{span} \left\{ I^{\alpha_1} J_2^{\alpha_2} J_3^{\alpha_3}
    \ldots J_d^{\alpha_d} \heff^j  : |\alpha|+j=n/2 \right\} \, ,
\label{2-15.2}
\end{equation}
where $I = q_1 p_1$ and $J_k = (q_k^2 + p_k^2)/2$, with $k = 2, \ldots, d$,
are the analogues of the classical integrals.

Applying the transformation (\ref{2-10}), with the generating
function defined by Eq.~(\ref{2-15}), for $n=3,\ldots,N$, and truncating the
resulting Taylor series (\ref{2-03}) at the $N^\mathrm{th}$ order one arrives at the
Weyl symbol $H_{\mathrm{QNF}}^{(N)}$ corresponding to the $N^{\mathrm{th}}$
order  {\it quantum normal form} (QNF) of the Hamiltonian
$\hat{H}$:
\begin{equation}
  H_{\mathrm{QNF}}^{(N)}(\vecz;\heff) = E_0 + \sum_{s=2}^N H_s^{(N)}(\vecz;\heff) \, .
\label{2-15.5}
\end{equation}
The $N^{\mathrm{th}}$ order QNF operator $\hat{H}_{\mathrm{QNF}}^{(N)}$ is
then given by
\begin{equation}
  \hat{H}_{\mathrm{QNF}}^{(N)} = \mathrm{Op}\left[ H_{\mathrm{QNF}}^{(N)}
  \right] \, ,
\label{2-16}
\end{equation}
where $ \mathrm{Op}\left[ \cdot \right]$ is the Weyl map defined in \eqref{eq:def_Weyl_map}.
\rem{ using the Dirac notations,
\begin{eqnarray}
  \mathrm{Op}[F] &=& \iint \frac{d\vecq d\vecp}{\left( 2\pi\heff \right)^d}
  F(\vecq,\vecp;\heff) \nonumber \\ && \times \displaystyle \int
  d\vecx | \vecq-\vecx/2 \rangle e^{-i\vecp\vecx / \heff}
  \langle \vecq+\vecx/2 |
\label{2-17}
\end{eqnarray}
defines the Weyl quantization of a symbol $F(\vecq,\vecp;\heff)$.} 
The Weyl  quantization of the classical integrals $I$ and $J_k$, $k=2,\ldots,d$, are
\begin{eqnarray}
  \hat{I} &\equiv& \mathrm{Op}[I] = \displaystyle \frac{1}{2} (\hat{q} \hat{p}
  + \hat{p} \hat{q}) \, , \label{2-20a}\\
  \hat{J}_k &\equiv& \mathrm{Op}[J_k] = \displaystyle \frac{1}{2} (\hat{q}_k^2 +
  \hat{p}_k^2),\quad k=2,\ldots,d.
  \label{2-20b}
\end{eqnarray}
Using Eq.~(\ref{2-05}) and the linearity of the Weyl quantization we get
\begin{equation}
  \hat{H}_2^{(2)} = \lambda \hat{I} + \sum_{k=2}^d \omega_k \hat{J}_k  \, .
\label{2-19}
\end{equation}
Since the higher order terms in \eqref{2-15.5} are polynomials in $I$ and $J_k$, $k=2,\ldots,d$ (see \eqref{2-15.2}), 
we need to know how to quantize powers of $I$ and $J_k$.  As shown in \cite{wsw08}
this can be accomplished using the recurrence relations
\begin{equation}
\mathrm{Op}\left[ I^{n+1} \right] = \hat{I}    \mathrm{Op}\left[ I^{n} \right] - \left(\frac{\hbar}{2} \right)^2 n^2  \mathrm{Op}\left[ I^{n-1} \right]
\label{eq:recurrence_I}
\end{equation}
and
\begin{equation}
\mathrm{Op}\left[ J_k^{n+1} \right] = \hat{J}_k    \mathrm{Op}\left[ J_k^{n} \right] + \left(\frac{\hbar}{2} \right)^2 n^2  \mathrm{Op}\left[ J_k^{n-1} \right]
\label{eq:recurrence_J}
\end{equation}
for $k=2,\ldots,d$. Hence, $ \hat{H}_{\mathrm{QNF}}^{(N)}$ is a polynomial function of the operators $\hat{I}$ and $\hat{J}_k$:
\begin{eqnarray}
  \hat{H}_{\mathrm{QNF}}^{(N)} &=& K_{\mathrm{QNF}}^{(N)}(\hat{I}, \hat{J}_2,
  \hat{J}_3, \ldots, \hat{J}_d) \nonumber\\
  &=& E_0 + \lambda \hat{I} + \sum_{k=2}^d \omega_k \hat{J}_k 
  \nonumber\\
  && \!\! + \! \sum_{n=2}^{\lfloor N/2 \rfloor} \! \sum_{|\alpha|+j = n} \!\!\!\! k_{n,\alpha,j}
  \hat{I}^{\alpha_1} \hat{J}_2^{\alpha_2} \ldots \hat{J}_d^{\alpha_d} \heff^j
  \, . \;\;\;
\label{2-21}
\end{eqnarray}
The coefficients $k_{n,\alpha,j}$ are systematically obtained by the
QNF procedure to compute the symbol $H_{\mathrm{QNF}}^{(N)} $ as
desribed above and the recurrence relations \eqref{eq:recurrence_I}
and \eqref{eq:recurrence_J}.  So the full procedure to compute
$\hat{H}_{\mathrm{QNF}}^{(N)} $ is algebraic in nature, and can be
implemented on a computer. Our software for computing the quantum
normal form as well as the classical normal form which is recovered
for $\heff=0$ is publicly available at
\url{http://lacms.maths.bris.ac.uk/publications/software/index.html}.

We stress that $\hat{H}_{\mathrm{QNF}}^{(N)}$ represents an
$N^\mathrm{th}$ order approximation of the operator obtained from
conjugating the original Hamiltonian $\hat{H}$ by the unitary
transformation
\begin{equation}
\hat{U} =  \ue^{-\ui\hat{W}_1/\heff}  \ue^{-\ui\hat{W}_2/\heff}   \cdots   \ue^{-\ui\hat{W}_N/\heff} \,,
\label{eq:def_U}
\end{equation} 
where we used the fact that the first two steps in the sequence
\eqref{eq:seq_trafos} can also be implemented using suitable
generators $\hat{W}_1$ and $\hat{W}_2$ (see \cite{wsw08} for
more details).
This  is why it is legitimate to use
$\hat{H}_{\mathrm{QNF}}$ instead of $\hat{H}$ in analyzing such properties of
the system as the CRP.

\rem{
It  follows from the construction of the QNF that $\mathcal{D}
H_{\mathrm{QNF}}^{(N)} = 0$ and, therefore,
\begin{equation}
  [ \hat{H}_2^{(2)} , \hat{H}_{\mathrm{QNF}}^{(N)} ] = 0 \, ,
\label{2-18}
\end{equation}
where $\hat{H}_2^{(2)} = \mathrm{Op}[H_2^{(2)}]$. According to
Eq.~(\ref{2-05}) $H_2^{(2)} = \lambda I + \sum_{k=2}^d \omega_k J_k $,
where
\begin{eqnarray}
  \hat{I} &\equiv& \mathrm{Op}[I] = \displaystyle \frac{1}{2} (\hat{q} \hat{p}
  + \hat{p} \hat{q}) \, , \label{2-20a}\\
  \hat{J}_k &\equiv& \mathrm{Op}[J_k] = \displaystyle \frac{1}{2} (\hat{q}_k^2 +
  \hat{p}_k^2),\quad k=2,\ldots,d,
  \label{2-20b}
\end{eqnarray}
are the  operators associated with the classical integrals. Therefore, it
follows from Eq.~(\ref{2-15.5}) together with Eqs.~(\ref{2-12}),
(\ref{2-15.2}), and (\ref{2-19}) that
} 

The main advantage of having the Hamiltonian in the form of a
polynomial in the operators $\hat{I}$ and $\hat{J}_k$, $k=2,\ldots,d$,
is that the eigenstates of the QNF operator
$\hat{H}_{\mathrm{QNF}}^{(N)}$ can be chosen to be simultaneously the
eigenstates of the operators $\hat{I}$ and $\hat{J}_k$, whose spectral
properties are well known:
\begin{equation}
  \hat{H}_{\mathrm{QNF}}^{(N)} | I,n_2,\ldots,n_d \rangle = E |
  I,n_2,\ldots,n_d \rangle \, ,
\label{2-22}
\end{equation}
where
\begin{eqnarray}
  \hat{I} | I,n_2,\ldots,n_d \rangle &=& I | I,n_2,\ldots,n_d \rangle \, ,
  \label{2-23} \\
  \hat{J}_k | I,n_2,\ldots,n_d \rangle &=& \heff (n_k+1/2) | I,n_2,\ldots,n_d
  \rangle \;\;\;\;\;
  \label{2-24}
\end{eqnarray}
with $n_k \in \mathbb{N}_0$ and $k=2,\ldots,d$, and the energy being given by
\begin{equation}
  E = K_{\mathrm{QNF}}^{(N)} \left( I, \heff (n_2+1/2), \ldots, \heff
    (n_d+1/2) \right) \, .
\label{2-25}
\end{equation}
Effectively, the QNF procedure yields an approximation of the
original Hamiltonian, $\hat{H}$, in terms of the operator
$\hat{H}_{\mathrm{QNF}}^{(N)}$ whose classical counterpart is
integrable while the classical counterpart of $\hat{H}$ is in general
not integrable.  The approximation is only valid in the neighborhood
of the saddle equilibrium point. However, it is crucial to note that
this local approximation is sufficient to compute the cumulative
reaction probability which in terms of the QNF is given by
\cite{seideman,wsw08}
\begin{equation}
  \crp(E) = \!\! \sum_{n_2,\ldots,n_d} \! \left[ 1+\exp\left( -2\pi
      \frac{I(E,n_2,\ldots,n_d)}{\heff} \right) \right]^{-1} \!\! ,
\label{2-26}
\end{equation}
where the summation runs over all $n_2,\ldots,n_d$, and for given
energy $E$ and quantum numbers $n_2,\ldots,n_d$, the quantity $I$ in
\eqref{2-26} is implicitly defined by Eq.~(\ref{2-25}).


\section{Collinear hydrogen- and nitrogen-exchange reactions}
\label{sec_reactions}

In this section we demonstrate the efficiency and the capability of the QNF theory
by applying it to the computation of the CRP for
collinear triatomic reactions. To this end we focus on Hamiltonians of the
form
\begin{equation}
  \hat{H} \equiv H(\hat{q}_1, \hat{q}_2, \hat{p}_1, \hat{p}_2)
  = \frac{1}{2} \left( \hat{p}_1^2 + \hat{p}_2^2 \right) + V(\hat{q}_1, \hat{q}_2) \, ,
\label{3-01}
\end{equation}
where $V(q_1,q_2)$ gives the Born-Oppenheimer potential energy surface
(PES) of a two-dimensional atomic system. Here, $q_1$ and $q_2$ are
the Delves mass-scaled coordinates \cite{delves}, and the effective
Planck's constant is given by $\heff = \mu^{-1/2}$, where $\mu$ is the
(dimensionless) reduced mass of the triatomic system (note that the
electronic mass is 1 in the atomic units we are using).

The PES is assumed to possess a single saddle point governing the reaction from
the asymptotic reactants and products  states. In this paper we analyze the following
collinear exchange reactions:
\begin{eqnarray}
\mathrm{H} + \mathrm{H}_2 &\rightarrow& \mathrm{H}_2 + \mathrm{H}\,,
\label{3-02}\\
\mathrm{N} + \mathrm{N}_2 &\rightarrow& \mathrm{N}_2 + \mathrm{N}\,,
\label{3-03}
\end{eqnarray}
where various isotopes of hydrogen are considered. The Porter-Karplus (PK)
PES \cite{porter} is taken to model the hydrogen exchange reaction
(\ref{3-02}), and the London-Eyring-Polanyi-Sato (LEPS) PES \cite{lagana}
is adopted for the nitrogen exchange reaction (\ref{3-03}).

We applied the algorithm presented in Sec.~\ref{sec_theory} to construct the
QNF Hamiltonian of various orders for the triatomic systems in
(\ref{3-02}) and (\ref{3-03}). Then, the QNF Hamiltonian was used to
compute the CRP for a range of reaction energies $E$ in accordance with
Eq.~(\ref{2-26}). The obtained CRP-vs-energy curves, $\crp(E)$, were later
compared to the results of the full {\it reactive quantum scattering}
calculations \cite{hauke,kupperman}. The latter were performed by integrating
the coupled multichannel Schr\"odinger equation in hyperspherical coordinates
\cite{hauke,kupperman} from the strong interaction region to the asymptotic
reactant and product configurations. The log-derivative matrix method of
Manolopoulos and Gray \cite{manol} together with the six-step symplectic
integrator of McLachlan and Atela \cite{mclach} was used to integrate the
radial Schr\"odinger equation.

\begin{figure}[h]
\centerline{\epsfig{figure=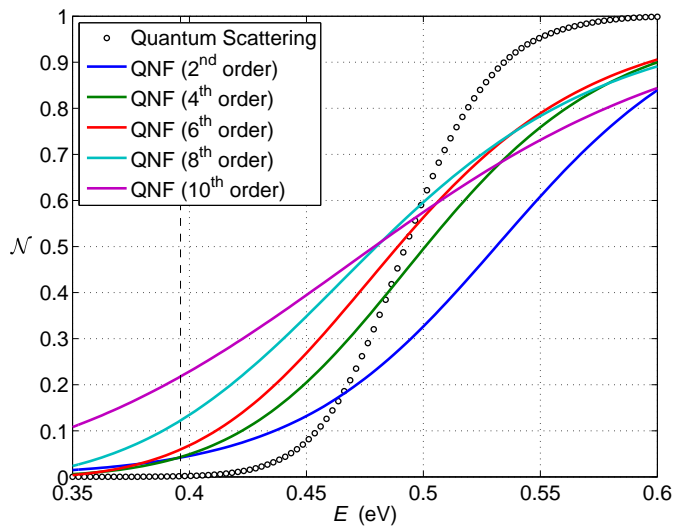,width=3.5in}}
\caption{Cumulative reaction probability as a function of the total
  energy, $\crp(E)$, for the collinear reaction (\ref{3-02})
  involving three $^1$H atoms. The effective Planck's constant is $\heff
  \approx 3.07 \times 10^{-2}$. The vertical dashed line shows the
  saddle point energy, $E_0$, of the PK PES.}
\label{fig-1}
\end{figure}

Figure~\ref{fig-1} shows the CRP, $\crp(E)$, as a function of the total energy
$E$ for a collinear hydrogen, $^1$H, exchange reaction, Eq.~(\ref{3-02}), on
the PK PES. The circular points represent $\crp(E)$ obtained in the reactive
quantum scattering calculation, and can, therefore, be regarded as the `exact'
CRP values. The vertical dashed line shows the saddle point energy, $E_0$, of
the PK PES. The five solid colored lines represent the $\crp(E)$ curves
corresponding to different orders, $N=2,4,\ldots,10$, of the QNF computation.
As we argue in Sec.~\ref{sec_convergence}, one of the sources of the apparent
failure of the QNF method to reproduce the correct values of the CPR in the
collinear $^1$H triatomic system is the very slow convergence (or perhaps even
divergence) of the QNF expansion for the value of the effective Planck's
constant, $\heff \approx 3.07 \times 10^{-2}$, characterizing this particular
reacting system. Another reason for the QNF theory to be unable to predict
correct CPR values for the hydrogen exchange reaction is the importance of the
corner cutting {\it tunneling trajectories} \cite{miller86} in reaction dynamics of
light-atom systems. These  tunneling trajectories avoid passing through the
immediate neighborhood of the saddle-center-\ldots-center equilibrium point in
phase space and, therefore, their contribution to the CRP can not be captured
by the QNF theory.

\begin{figure}[h]
\centerline{\epsfig{figure=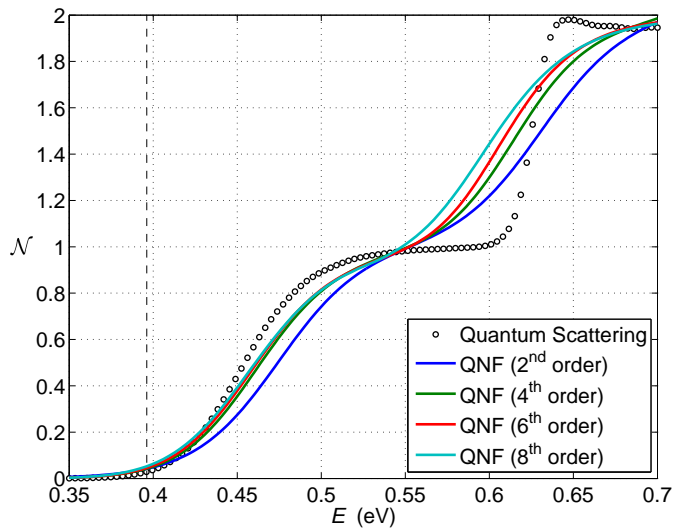,width=3.5in}}
\caption{Cumulative reaction probability as a function of the total energy,
  $\crp(E)$, for the collinear reaction (\ref{3-02}) with $^{3}$H (tritium) isotopes
  of hydrogen. The effective Planck's constant is $\heff \approx
  1.77 \times 10^{-2}$. The vertical dashed line shows the saddle point
  energy, $E_0$, of the PK PES.}
\label{fig-2}
\end{figure}

Figure~\ref{fig-2} presents the CRP-vs-energy curves obtained in the reactive
quantum scattering approach (circular points) and by the QNF calculation
(colored solid lines) of different orders, $N=2,4,\ldots,8$, for the triatomic
collinear system of $^3$H (tritium) isotopes of hydrogen. The vertical
dashed line shows the saddle point energy, $E_0$, of the PK PES. The effective
Planck's constant characterizing the system is now $\heff \approx 1.77 \times
10^{-2}$. The convergence of the QNF $\heff$-expansion, for the energies up to
$\sim$0.54 eV, is now evident from the figure. However, the QNF-predicted CRP
values approximate the reactive quantum scattering $\crp(E)$ data
only at small energies. As in the case of the $^1$H exchange reaction, see
fig.~\ref{fig-1}, we attribute the disagreement of the QNF and reactive
quantum scattering CRP values to the non-negligible contributions of 
tunneling trajectories which avoid passing through the neighborhood of the 
saddle. 

\begin{figure}[h]
\centerline{\epsfig{figure=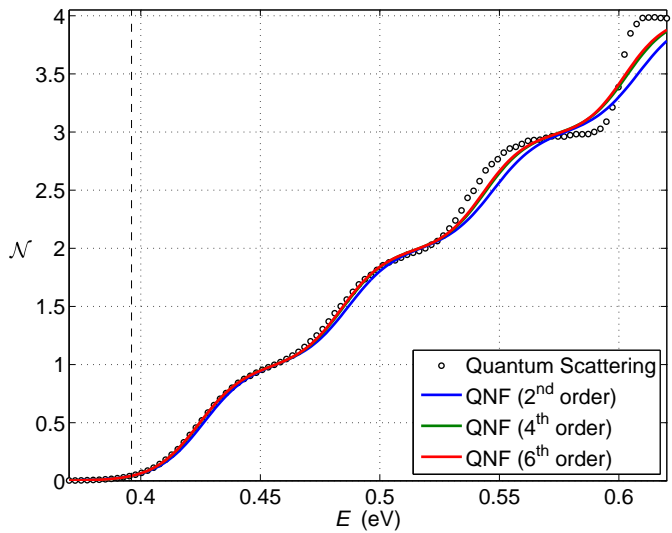,width=3.5in}}
\caption{Cumulative reaction probability as a function of the total energy,
  $\crp(E)$, for the collinear reaction  (\ref{3-02}) with hypothetical
  $^{20}$H isotopes of hydrogen. The effective Planck's constant is $\heff
  \approx 6.9 \times 10^{-3}$. The $\crp(E)$ curves obtained with the
  $4^{\mathrm{th}}$ and $6^{\mathrm{th}}$ order QNF are basically
  indistinguishable for most of the energy range. The vertical dashed line
  shows the saddle point energy, $E_0$, of the PK PES.}
\label{fig-3}
\end{figure}

Figure~\ref{fig-3} presents the results of the CRP calculations for a
collinear system of three hypothetical $^{20}$H isotopes of hydrogen. As
before, the circular data points correspond to the reactive quantum scattering
data and are treated as exact CRP values. The three colored solid lines show the
QNF $\crp(E)$ curves of orders $N=2,4,6$; the $\crp(E)$ curves obtained with
the $4^{\mathrm{th}}$ and $6^{\mathrm{th}}$ order QNF are essentially
indistinguishable for most of the energy range. The vertical dashed line shows
the saddle point energy, $E_0$, of the PK PES. The model system is
characterized by $\heff \approx 6.9 \times 10^{-3}$. The convergence of the
QNF $\heff$-expansion, as well as the quantitative agreement of the QNF
predictions and exact CPR values for energies $E \lesssim 0.45$ eV, is
evident from the figure.

Comparison of figs.~\ref{fig-1}-\ref{fig-3} allows us to conclude that, while
basically failing for systems of light atoms, the QNF method of computing the
CPR proves very effective for treating heavy-atom reactive systems. On the
contrary, the full reactive quantum scattering computations are only feasible
for reactive systems consisting of light atoms, and the computations rapidly
become formidable as the atomic mass is increased \cite{walker}.

\begin{figure}[h]
\centerline{\epsfig{figure=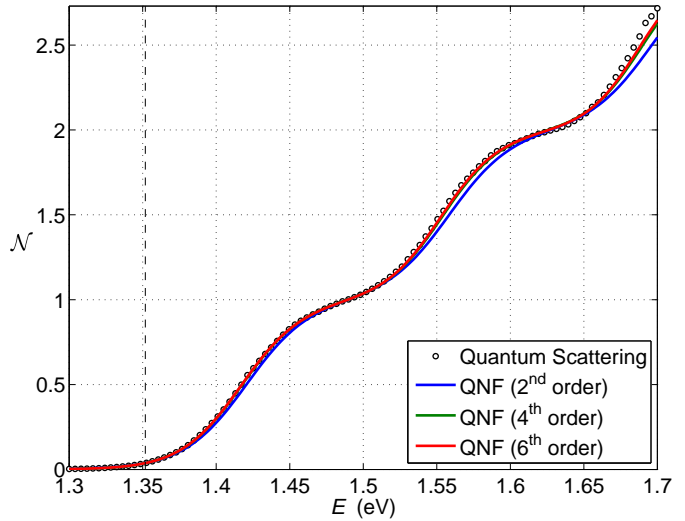,width=3.5in}}
\caption{Cumulative reaction probability as a function of the total energy,
  $\crp(E)$, for the collinear nitrogen exchange reaction (\ref{3-03}). 
  The effective Planck's constant is $\heff \approx 8.2 \times
  10^{-3}$. The $\crp(E)$ curves obtained with the $4^{\mathrm{th}}$ and
  $6^{\mathrm{th}}$ order QNF are essentially indistinguishable for most of the
  energy range. The vertical dashed line shows the saddle point energy, $E_0$,
  of the LEPS PES.}
\label{fig-4}
\end{figure}

Finally, in order to further illustrate the efficiency of the QNF technique
for treating heavy-atom systems we compute the CRP for the collinear nitrogen
exchange reaction \eqref{3-03} on the LEPS PES.  Figure~\ref{fig-4} compares the CRP values
obtained in the reactive quantum scattering calculation (circular data points)
and those given by the QNF analysis (colored solid lines) of orders $N=2,4,6$.
The system is characterized by $\heff \approx 8.2 \times 10^{-3}$. The
vertical dashed line shows the saddle point energy, $E_0$, of the LEPS PES.
The $\crp(E)$ curves obtained with the $4^{\mathrm{th}}$ and $6^{\mathrm{th}}$
order QNF are essentially indistinguishable for most of the energy range; this fact
signals the rapid convergence of the QNF $\heff$-expansion for the given value
of the effective Planck's constant. The quantitative agreement of the exact
and QNF values of $\crp(E)$ extends up to energies of $\sim$1.5 eV.

The QNF calculation of the CRP requires significantly less
computational time than the corresponding full quantum reactive
scattering calculation. For example, the 6$^\text{th}$ order QNF computation of
the nitrogen-exchange CRP curve in Fig.~\ref{fig-4} took about 10
minutes on a 2.6 GHz processor, 2 GB RAM computer, while the
corresponding full quantum reactive scattering computation took more
than 12 hours on the same machine. The QNF approach becomes even more
advantageous for treating chemical systems of atoms heavier than
nitrogen: the expense of the full quantum computations rapidly grows
with the number of asymptotic channels (and, therefore, with mass)
\cite{walker}, while the QNF expansion only becomes more rapidly
convergent making the corresponding analysis computationally cheaper.


\section{Convergence of QNF}
\label{sec_convergence}

While it is well known that for $d=2$ degrees of freedom, the
classical normal form (CNF) converges in the neighborhood of
saddle-center equilibrium points (see, e.g., \cite{giorgilli, moser} )
this is not clear for the QNF (for the first results in this direction
see \cite{anikin}). Still, in the following we provide a qualitative
discussion of the convergence of the QNF based on our calculations
performed for the triatomic collinear reactions of
Sec.~\ref{sec_reactions}.

The QNF approximates the Hamiltonian of the reaction system in a phase-space
vicinity of the saddle-center equilibrium point. Thus, for instance, in
computing the CRP one only expects this approximation to render reliable
results in a certain energy range around the saddle point energy $E_0$ of the
PES under consideration. The energy difference $(E-E_0)$ may therefore be
considered as one small parameter in the QNF expansion. The role of the other
small parameter is played by the effective Planck's constant, $\heff$. It is
the convergence of the QNF with respect to this second small parameter that we
focus on in this section.

We proceed by considering the right hand side of Eq.~(\ref{2-25}), i.e., the
QNF, at $I=0$, corresponding to no `energy' in the reaction coordinate, and $n_2=0$, giving the zero-point
`vibrational energy' of the transverse degree of freedom. Then, Eq.~(\ref{2-25}) becomes
\begin{equation}
  E = E_0 + \sum_{n=1}^{\lfloor N/2 \rfloor} c_n \heff^n \, .
\label{4-01}
\end{equation}
For the case of the PK PES the first five expansion coefficients are $c_1 =
0.161982$, $c_2 = 1.193254$, $c_3 = 14.90023$, $c_4 = 378.7950$, and $c_5 =
1227.035$. As $N \rightarrow \infty$ the radius of convergence $\heff^{(0)}$
of the sum in Eq.~(\ref{4-01}) is given by
\begin{equation}
  \heff^{(0)} = \lim_{n\to \infty} \frac{ c_n}{c_{n+1} } \, .
\label{4-02}
\end{equation}
Here, we make a crude estimate of $\heff^{(0)}$ by only considering the first
five expansion coefficients in Eq.~(\ref{4-02}), i.e., $c_n$ with
$n=1,\ldots,5$; then, the radius of convergence is given by $\heff^{(0)} \sim 0.04$.

The estimated value of $\heff^{(0)}$ sheds light on the seeming inefficiency
of the QNF theory for CRP computations in light atom reactions. Indeed, the
$^1$H exchange reaction, see Fig.~\ref{fig-1}, is characterized by $\heff =
3.07 \times 10^{-2}$. This value being close to $\heff^{(0)}$ signals that the
corresponding QNF expansion converges very slowly, if at all, and, possibly,
terms of orders far beyond $N=10$ are needed for a reliable CRP prediction in
Fig.~\ref{fig-1}.

In the case of the $^3$H exchange reaction the effective Planck's constant is
$\heff = 1.77 \times 10^{-2}$ and is thus smaller than $\heff^{(0)}$. This fact is in
agreement with the apparent speed-up of the convergence of the CRP values, see
Fig.~\ref{fig-2}, in comparison with the $^1$H case. Finally, the convergence
is very fast and pronounced for the case of the heavy (hypothetical) $^{20}$H
atoms, see Fig.~\ref{fig-3}, for which $\heff = 6.9 \times 10^{-3}$ which is
much smaller that the estimated convergence radius.

\section{Conclusions}
\label{sec_conclusions}

In this paper we used the quantum normal form (QNF) approach to
quantum transition state theory \cite{schubert_prl,wsw08} for computing the cumulative reaction
probability for triatomic collinear reactions. 
The QNF leads to a realization  of quantum transition state theory  which is very much in the spirit of (classical) transition state theory. 
Similar to the classical case where a recrossing free dividing surface can be constructed from a classical normal form such that reaction probabilities can be computed from the flux through the dividing surface, the QNF can be viewed to give quantum reaction probabilities as the quantum  mechanical flux through the same (classically recrossing free) dividing surface. So unlike reactive scattering techniques which involve full, global quantum computations, the  QNF realization of quantum transition state theory requires only local information in the neighborhood of the saddle equilibrium point which governs the reaction. In this paper we demonstrated, that for heavy atom systems (comprised of ten or more nucleons) the QNF this way indeed gives a very efficient method for computing cumulative reaction probabilities. 
Here  we measure `efficiency'   by the effort for both implementing and computing the QNF. The latter are both comparable to implementing and computing the classical normal form which lead to the realization of classical transition state theory (in particular for multidimensional systems). The major difference between the classical and quantum case is that the QNF computation involves the Moyal bracket which slightly more complicated (and thus computationally more expensive)  than the Poisson bracket in the classical case. Nevertheless the efforts for implementing and computing the QNF are far lower than for the full reactive scattering computations to which we compared our results. 

We saw, however, that for reactions involving light reactions (such as the hydrogen exchange reaction) the QNF gave only very poor results. We attributed the failure of the QNF computation in these cases to the presence of corner cutting tunneling trajectories which are not captured by the QNF.  This way the QNF and reactive scattering methods can be viewed as complementary methods where the latter gives very good results for light atom systems and the former displays its full power especially for heavy atom systems for which reactive scattering approaches become very difficult or even unfeasible due  to the growing number of reactive channels that have to be taken into account \cite{walker}. 

We note that also other approximation techniques such as the
initial value representation (IVR) \cite{ivr} have been shown to be
fruitful for reaction probability analysis of collinear triatomic
reactions \cite{garashchuk}. However, in order to properly account for
interference effects the IVR method requires propagation of a huge
number of classical trajectories and, therefore, can pose difficulties
for application to high-dimensional atomic systems whereas the difficulties in computing the QNF do not grow so rapidly with the number of degrees of freedom. In fact it would be very interesting to make a detailed comparison between the QNF and the IVR approach.

Another benefit of the QNF approach to compute cumulative reaction probabilities lies in the fact that it involves only little (local) information of the Born-Oppenheimer PES; namely the Taylor expansion of the PES about the saddle equilibrium point governing the reaction. In fact we saw that highly accurate results over quite a broad  energy range can already be obtained from the 4$^{\text{th}}$ or 6$^{\text{th}}$ Taylor expansion which enters the QNF of the same order. This is especially useful for systems for which the computation of the global PES required in other methods is very difficult.

\rem{
Unlike the reactive quantum scattering technique for which the  computational
expenses rapidly grow with the number of asymptotic channels
rendering the method extremely computationally intensive for analyzing
systems of heavy atoms \cite{walker}, the QNF theory only deals with a
small neighborhood of the PES saddle point only and this way, provides an
efficient, computationally inexpensive technique for CRP calculations
in heavy atom reactions. Other approximation techniques such as the
initial value representation (IVR) \cite{ivr} have been shown to be
fruitful for reaction probability analysis of collinear triatomic
reactions \cite{garashchuk}. However, in order to properly account for
interference effects the IVR method requires propagation of a huge
number of classical trajectories and, therefore, can pose difficulties
for application to high-dimensional atomic systems. A direct
comparison between the QNF approach and the IVR approach would be very
interesting.

Another significant advantage of the QNF approach over the full
reactive quantum scattering and semiclassical methods like IVR is that the QNF
computations require one to know the
Born-Oppenheimer PES only in a vicinity of the saddle (the first few Taylor expansion coefficients are sufficient) rather than the full PES. 
This may render the
QNF approach the preferred computational method for chemical systems
for which determination of the complete PES is especially complicated
or simply unfeasible.
} 


\section{Acknowledgments} 

A.G. and H.W. acknowledge support by EPSRC under grant number
EP/E024629/1. Part of this work was carried out using the
computational facilities of the Advanced Computing Research Centre,
University of Bristol. S.W. acknowledges the support of ONR Grant
No.~N00014-01-1-0769, and also the stimulating environment of the NSF
sponsored Institute for Mathematics and its Applications (IMA) at the
University of Minnesota, where some of this work was carried out. We
are also grateful to Prof. Gregory S. Ezra for reading an earlier
version of this manuscript and offering useful comments.



\begin{thebibliography}{99}

\bibitem{PechukasMcLafferty73}
P.~Pechukas and F.~J. McLafferty, J. Chem. Phys. {\bf 58}, 1622
  (1973).
  
\bibitem{PechukasPollak78} P. ~Pechukas and E.~ Pollak,
  J. Chem. Phys. {\bf 69}, 1218 (1978).


\bibitem{Wiggins94}
Wiggins S.:  Normally Hyperbolic Invariant Manifolds in Dynamical
  Systems.  Springer: Berlin 1994
  
\bibitem{UJPYW01} T.~Uzer, C.~Jaff{\'e}, J.~Palaci{\'a}n, P.~Yanguas,
  and S.~Wiggins, Nonlinearity, {\bf 15}, 957 (2001).
  
  \bibitem{WaalkensWiggins04}
H.~Waalkens and S.~Wiggins, 
  J. Phys. A, {\bf 37, L435, } (2004).
  
  
\bibitem{schubert_prl} R.~Schubert, H.~Waalkens, and S.~Wiggins, Phys. Rev.
  Lett. {\bf 96}, 218302 (2006).
  
  
  
\bibitem{wsw08} H.~Waalkens, R.~Schubert, and S.~Wiggins, Nonlinearity
  {\bf 21}, R1 (2008).

  
  \bibitem{Miller98} Miller, W. H.: J. Phys.
  Chem. A,  {\bf 102},  793 (1998)
  
\bibitem{Miller1} Miller, W. H.: \emph{Spiers {M}emorial {L}ecture},
  Farad. Discuss. {\bf 110} 1 (1998)
  
  

\bibitem{seideman} T.~Seideman and W.~H.~Miller, J. Chem. Phys. {\bf
    95}, 1768 (1991).

\bibitem{miller98} W.~H.~Miller, J. Phys. Chem. A {\bf 102}, 793 (1998).













\bibitem{delves} L.~M.~Delves, Nucl. Phys. {\bf 9}, 391 (1959); Nucl. Phys.
  {\bf 20}, 275 (1960).





\bibitem{porter} R.~N.~Porter and M.~Karplus, J. Chem. Phys. {\bf 40}, 1105
  (1964).

\bibitem{lagana} A.~Lagana, E.~Garcia, and L.~Ciccarelli, J. Phys. Chem. {\bf
    91}, 312 (1987).





\bibitem{hauke} G.~Hauke, J.~Manz, and J.~R\"omelt, J. Chem. Phys. {\bf 73},
  5040 (1980).

\bibitem{kupperman} A.~Kuppermann, J.~A.~Kaye, and J.~P.~Dwyer, Chem. Phys.
  Lett. {\bf 74}, 257 (1980).

\bibitem{manol} D.~E.~Manolopoulos and S.~K.~Gray, J. Chem. Phys. {\bf 102},
  9214 (1995).

\bibitem{mclach} R.~I.~McLachlan and P.~Atela, Nonlinearity {\bf 5}, 541
  (1991).




\bibitem{miller86} W.~H.~Miller, Science {\bf 233}, 171 (1986).




\bibitem{walker} R.~B.~Walker and J.~C.~Light,
  Ann. Rev. Phys. Chem. {\bf 31}, 401 (1980).






\bibitem{giorgilli} A.~Giorgilli, Discr. Cont. Dyn. Sys., {\bf 7}(4),
  855 (2001).

\bibitem{moser} J.~Moser, Comm. Pure Appl. Math., {\bf 11}, 257
  (1958).





\bibitem{anikin} A.~Anikin, Reg. Chaot.  Dyn. {\bf 13}, 377 (2008).





\bibitem{ivr} W.~H.~Miller, J. Phys. Chem. A {\bf 105}, 2942 (2001); Y.~Elran
  and K.~G.~Kay, J. Chem. Phys. {\bf 114}, 4362 (2001); J. Chem. Phys {\bf
    116}, 10577 (2002).

\bibitem{garashchuk} S.~Garashchuk, F.~Grossmann, and D.~Tannor, J. Chem.
  Soc., Faraday Trans. {\bf 93}, 781 (1997).






\end{thebibliography}
\end{document}